\documentclass[tightenlines,a4paper,fleqn,altaffilletter,twoside,twocolumn,
               floatfix,nofootinbib]{revtex4}

\usepackage[english]{babel}
\usepackage{amsmath,amssymb}
\usepackage{graphicx}
\usepackage{url}

\def\om{{\omega}}
\def\eps{{\epsilon}}
\renewcommand{\vec}[1]{{\mathbf{#1}}}

\begin{document}

\title{Why a splitting in the final state cannot explain the GSI-Oscillations}
\author{Alexander Merle} \email[Email: ]{Alexander.Merle@mpi-hd.mpg.de}
\affiliation{Max--Planck--Institut f\"ur Kernphysik\\
             Postfach 10 39 80, D--69029 Heidelberg, Germany}

\begin{abstract}
In this paper, I give a pedagogical discussion of the GSI anomaly. Using two different formulations, namely the intuitive Quantum Field Theory language of the second quantized picture as well as the language of amplitudes, I clear up the analogies and differences between the GSI anomaly and other processes (the Double Slit experiment using photons, $e^+ e^- \rightarrow \mu^+ \mu^-$ scattering, and charged pion decay). In both formulations, the conclusion is reached that the decay rate measured at GSI cannot oscillate if only Standard Model physics is involved and the initial hydrogen-like ion is no coherent superposition of more than one state (in case there is no new, yet unknown, mechanism at work). Furthermore, a discussion of the Quantum Beat phenomenon will be given, which is often assumed to be able to cause the observed oscillations. This is, however, not possible for a splitting in the final state only.\\

\end{abstract}

\maketitle

\section{\label{sec:intro} Introduction}

In the last months, a measurement of the lifetime of several highly charged ions with respect to electron capture (EC) decays at GSI Darmstadt~\cite{Litvinov:2008rk} has caused a lot of discussion: Instead of seeing only the exponential decay law, a superimposed oscillation has been observed. The
cause of this phenomenon, often referred to as {\it Darmstadt oscillations} or
{\it GSI anomaly}, is not yet clear and a huge debate arose whether it could be
related to neutrino
mixing~\cite{Ivanov:2008sd,Ivanov:2008nb,Ivanov:2008zzc,Faber:2008tu,Lipkin:2008ai,Ivanov:2008xw,Lipkin:2008in,Kleinert:2008ps,Walker:2008zzb,Lipkin:2009zy,Ivanov:2009rc,Ivanov:2009en,Ivanov:2009kt,Ivanov:2009ku,Lipkin:2009ge},
or
not~\cite{Kienert:2008nz,Giunti:2008ex,Giunti:2008im,Giunti:2008eb,Burkhardt:2008ek,Peshkin:2008vk,Peshkin:2008qz,Gal:2008sw,Cohen:2008qb,Giunti:2008db,Giunti:2009ds,Flambaum:2009di}.
Alternative attempts for an explanation involve spin-rotation
coupling~\cite{Lambiase:2008ki,Pavlichenkov:2008tm,Faber:2009mg,Faestermann:2009tj}, the interference of the final states~\cite{Isakov:2009yr}, or hyperfine excitation~\cite{Winckler:2009jm}. From the experimental side, two test-experiments (with, however, different systematics~\cite{Litvinov:2008hf}) have been performed~\cite{Vetter:2008ne,Faestermann:2008jt}.\\
Several times in this discussion, the analogy of the GSI-experiment to the famous historical Double Slit experiment using, e.g., photons~\cite{Young} has been drawn~\cite{Lipkin:2008ai,Giunti:2008im,Giunti:2008db}, which also led to lively debates at several meetings~\cite{meetings}. In this article, I show that the intuitive Quantum Field Theory (QFT) formulation of the problem always leads to the correct result. As there is still a lot of discussion in part of the community, it might be useful to give one more detailed explanation of the Quantum Mechanics (QM) involved. This can be done best by presenting easy and familiar examples that are not necessarily directly related to the GSI anomaly but do involve the same logical steps and are not under dispute. To do this, I start with the superposition principle and discuss the Double Slit experiment with photons, $e^+ e^- \rightarrow \mu^+ \mu^-$ scattering, and the experiment performed at GSI. Afterwards, the language of amplitudes is used to further justify the QFT-treatment by carefully considering several cases, where $\pi^+$ decay serves as additional example before the considerations are applied to the GSI-experiment, too. Furthermore, the so-called {\it Quantum Beats}~\cite{Chow:1975zz} will be discussed, a well-known phenomenon that could indeed cause oscillatory decay rates. This has been argued to cause the observation at GSI in several places (see, e.g., Refs.~\cite{Ivanov:2008sd,Ivanov:2009rc,Ivanov:2009en}). It will, however, be shown that this cannot cause the observed behavior if a splitting is only present in the final state.\\
In the course of the paper, we will see that all three languages naturally lead to the same result, namely that a splitting in the final state cannot explain the GSI anomaly. Depending on the field of the reader, one or the other part may be clearer, but in the end it turns out that the intuitive QFT picture is correct and in perfect agreement with the results obtained using probability amplitudes or the language of Quantum Beats, which are just different formulations of the same basic principles.

\section{\label{sec:disctinction} The Quantum Field Theory formulation of the problem}
\begin{table*}[t]
\centering
\begin{tabular}{|c||c|c|c|}\hline
Category & Double Slit & $e^+ e^-\rightarrow \mu^+ \mu^-$ & GSI experiment\\ \hline \hline
1  & No slit-monitoring at all & $e^+ e^-$-collider & N/A \\ \hline
2A & Monitoring \& read out  & N/A & GSI-like experiment with more kinematical accuracy\\ \hline
2B & Monitoring without read out & N/A & Actual GSI-experiment \\ \hline
\end{tabular}
\caption{\label{tab:class} The classification of the three examples.}
\end{table*}

The starting point for the discussion is the superposition principle in QM. One common formulation is~\cite{Peskin:1995ev}: ``When a process can happen in alternative ways, we {\it add} the amplitudes for each possible way.'' The problem in the interpretation arises in the term ``alternative ways'', because it is not a priori clear what the word ``way'' actually means, as well as in the word ``process'', which exhibits similar ambiguities.\\
Let us use the following terminology: {\it Process} means a reaction with a well-defined initial and final state, whereas {\it way} is a particular intermediate state of a process. E.g., the scattering reaction $e^+ e^- \rightarrow \mu^+ \mu^-$ is one single process, no matter by which way ($\gamma$-, $Z^0$-, or $H^0$-exchange at tree-level in the Standard Model (SM) of elementary particles) it is mediated. $Z^0 \rightarrow \nu_e \overline{\nu}_e$ and $Z^0 \rightarrow \nu_\mu \overline{\nu}_\mu$ are, however, two distinct processes.\\
Using this terminology, the superposition principle can be formulated in the following way:
\begin{enumerate}

\item If different ways lead from the same initial to the same final state in one particular process, then one has to add the respective partial amplitudes to obtain the total amplitude. The absolute square of this total amplitude is then proportional to the probability of the process to happen ({\it coherent summation}).

\item If a reaction leads to physically distinct final states, then one has to add the probabilities for the different processes ({\it incoherent summation}).

\end{enumerate}
If a certain situation belongs to category~1, an interference pattern will be visible (or oscillations, in case the interfering terms have different phases as functions of time), while if it belongs to category~2, there will be no interference. The remaining question is at which point the measurement comes in. This can be trivially said for point~2: Either the experimental apparatus is sufficiently good to distinguish between the different final states (2A) -- then no summation whatsoever is necessary simply because one can divide the data set into two (or more), one for each of the different final states. If this is not the case (2B), the experiment will be able to lead to either of the final states, but one would not know which one had been the actual result -- then one would simply have to add the probabilities for the different final states to occur in order to obtain the total probability.\\
What if we do such a measurement for category~1? If we can indeed distinguish several ways that a process can happen, then this has to be done by some measurement. Since this measurement then has selected one particular way, we have actually transformed a situation belonging to category~1 into a situation of category~2. However, then there would be no terms to interfere with -- the interference would have been ``killed''.\\
Let us now turn to Table~\ref{tab:class}, that illustrates how our three examples fit into the categories~1, 2A, and 2B. These three cases will be discussed one by one in the following.

\subsection{\label{sec:dslit} The Double Slit experiment with photons}

This is the ``classical'' situation of an experiment that reveals the nature of QM. It has first been performed by Thomas Young~\cite{Young} and has later been the major example to illustrate the laws of QM. Its basic procedure is the following: Light emitted coherently by some source (e.g.\ a laser) hits a wall with two slits, both with widths comparable to the wavelength of the light. If it hits a screen behind the wall, one will observe an interference pattern, as characteristic for wave-like objects (category~1). There is, however, the interpretation of light as photons, i.e., quanta of a well-defined energy. Naturally, one could ask which path such a photon has taken, meaning through which of the two slits it has travelled. The amazing observation is that, as soon as one can resolve this by monitoring the slits accurately enough, the interference pattern will vanish, no matter if one actually reads out the information of the monitoring (2A), or not (2B). The reason is that, regardless of using the information or not, the measurement itself has disturbed the QM process in a way that the interference pattern is destroyed~\cite{feynman}.\\
The key point is that one cannot even say that the photon takes only one way: In the QM-formulation, amplitudes are added (and not probabilities), and hence the photon does not take one way or the other (and we simply sum over the results), but it rather has a total amplitude that includes a partial amplitude to take way 1 as well as another partial amplitude way 2. By taking the absolute square of this sum of amplitudes, interference terms appear.\\
A QFT-formulation involving elementary fields only would be much more complicated: One would sum over the amplitudes for the photon to interact with each electron and each quark in the matter the slits are made of, after having propagated to this particular particle and before further propagating to a certain point on the screen. Of course, by using an effective formulation of the theory, one can find a much more economical description and the easiest one is to simply comprise all possible interactions into two amplitudes, one for going through the first and one for going through the second slit.\\
Let us go back to this effective formulation: If there is monitoring, one actually ``kills'' one of these two amplitudes, the other one remains, and the interference is destroyed. Whenever there has been such a measurement, the interference will vanishe. As we will see, the question is {\it if} in a certain situation a measurement has been performed (or is implicitly included in the process considered), no matter if the corresponding information is read out, or not.

\subsection{\label{sec:eemumu} $e^+ e^- \rightarrow \mu^+ \mu^-$ scattering at a collider}
\begin{figure*}[t]
  \begin{center}
    \includegraphics[height=2.3cm]{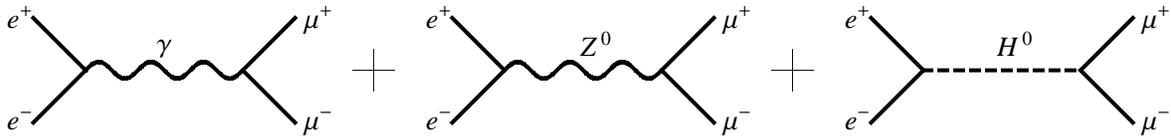}
  \end{center}
  \caption{The diagrams contributing to $e^+ e^-\rightarrow \mu^+ \mu^-$ in the SM.}
  \label{fig:eemm}
\end{figure*}

Let us now consider the scattering of $e^+ e^-$ to form a pair of muons. This is, differently from the Double Slit experiment, a fundamental process where only a small number of elementary particles is involved. If one wants to calculate the scattering probability, the amplitude for the process is again decisive. In the SM, there are only three possibilities for this process to happen at tree-level and in all three of them the $e^+ e^-$ pair annihilates to some intermediate (virtual) boson which in the end decays again, but this time into a $\mu^+ \mu^-$ pair. The intermediate particle can either be a photon, a $Z^0$-boson, or a Higgs scalar, see Fig.~\ref{fig:eemm}.\\
Here, we have three different ways to form the process. The difference to the Double Slit experiment, however, is that these three ways cannot be separated easily. In a real collider-experiment we are not able to say that the reaction $e^+ e^- \rightarrow \mu^+ \mu^-$ has taken place by the exchange of, e.g., a photon only, but it will always be the sum of the three diagrams (and a lot more, in case we include higher orders). Hence, this process will always fall into category 1 and interference terms will appear.\\
This is an easy and familiar example for the appearance of interference terms in a real experiment. In the next paragraph it will be shown exactly what is different in the case of the GSI-experiment.

\subsection{\label{sec:GSI} The GSI experiment}
\begin{figure}[t]
  \begin{center}
    \includegraphics[height=3cm]{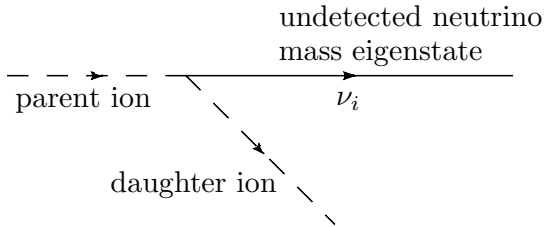}
  \end{center}
  \caption{The Feynman diagram for the GSI experiment.}
  \label{fig:feyn-GSI}
\end{figure}

The remaining question is what the situation looks like for the GSI-experiment. Even though the QFT-calculation of what happens is pretty straightforward, fitting everything in the language used above might be a bit more subtle. We will, however, see in Sec.~3.2 that the formulation in terms of amplitudes additionally justifies the result obtained here.\\
Let us at first consider the Feynman diagram of the process involved in the GSI-experiment in Fig.~\ref{fig:feyn-GSI}~\cite{Kienert:2008nz}: Here, in the absence of extreme kinematics (meaning that the $Q$-value of the reaction is large enough so that all neutrino mass eigenstates can be emitted), the neutrino is produced as electron neutrino. What happens to this neutrino? Since it is not detected, it escapes to infinity in the view of QFT (in the picture of second quantization). Physically, it loses its coherence after some propagation distance and travels as a unique mass eigenstate.\\
The key point is the following: Since the neutrino will not interact before it loses its coherence, it must be asymptotically a mass eigenstate. This can be shown easily: The coherence length of a (relativistic) neutrino is given by~\cite{Giunti:1991sx}
\begin{equation}
 L_{\rm coh}=2\sqrt{2} \sigma_x \cdot \frac{2p^2}{(\Delta m^2)_\odot},
 \label{eq:coh-length}
\end{equation}
where $\sigma_x$ is the size of the neutrino wave packet, $p$ is the momentum of the mean value neutrino momentum in the limit $m_\nu=0$, and $(\Delta m^2)_\odot=7.67\cdot 10^{-5}~{\rm eV}^2$~\cite{GonzalezGarcia:2007ib} is the solar neutrino mass square difference as known from neutrino oscillation experiments. The question is how to obtain an estimate for $\sigma_x$: If the nucleus was inside a lattice, one could estimate a width like the typical interatomic distance, $\sigma_x\sim 1$\AA{}, which would lead to $L_{\rm coh}\sim 2\cdot 10^8$~m. Of course, this precision cannot be reached in the GSI experiment. However, at least during the electron cooling~\cite{Steck}, the nucleus will be localized to some precision. Since the velocity of the nucleus is known, this information could in principle be extrapolated for each run. A fair estimate would then be the average distance between two electrons in the cooling process, which is roughly given by $1/\sqrt[3]{n}\sim 0.1$mm, where $n$ is the electron density~\cite{Bosch_ESR}. This leads to a more realistic coherence length of $L_{\rm coh}\sim 2\cdot 10^{14}$~m. The pessimistic case, where $\sigma_x$ is taken to be the approximate diameter $108.36~{\rm m}/\pi$~\cite{ESR} of the Experimental Storage Ring (ESR) produces $L_{\rm coh}\sim 6\cdot 10^{19}$~m. The mean free path of a neutrino in our galaxy, however, is roughly $1\cdot 10^{40}$~m (for an assumed matter density in the Milky Way of $1\cdot 10^{-23}~{\rm g/cm}^3$), so the assumption that the neutrino does not interact before losing its coherence is completely safe.\\
Even if we do not know in which of the three mass eigenstates the neutrino actually is, we know that it has to be in one of them. This knowledge is somehow obtained ``a posteriori'', since the mass eigenstate only reveals its identity after some propagation. But, by conservation of energy and momentum, one could treat the process as if the kinematical selection had already been present at the production point of the neutrino. This ``measurement'' is enforced by the physical conservation laws.\\
An analogous reasoning is given by Feynman and Hibbs~\cite{feynman}, using the example of neutron scattering: Neutrons prepared to have all spin up scatter on a crystal. If one of the scattered neutrons turns out to have spin down, one knows by angular momentum conservation that it must have been scattered by a certain nucleus. In principle, by noting down the spin state of every nucleus in the crystal before and after the measurement, one could find the corresponding scattering partner of the neutron without disturbing it. No matter if this would be difficult practically, by a physical conservation law one knows that a particular scattering must have been present, even if the corresponding nucleus is not ``read out''. Accordingly, the corresponding interference vanishes and the neutrons that have spin down after the scattering come out diffusely in all directions.\\
This can also be formulated in the language of wave packets: We have complete 4-momentum conservation for each single component (which is a plane wave!) of the wave packets, but if we consider the whole wave packet, its central momentum does not have to be conserved~\cite{Beuthe:2001rc,Giunti:2003ax}. However, all the different components can produce both possible neutrino mass eigenstates, but for a certain kinematical configuration of parent and daughter components only one of the mass eigenstates will actually be produced.\\
The rest is easy: If the GSI experiment had infinite kinematical precision, one could read out which of the mass eigenstates has been produced and it would clearly fall into category 2A. Since, however, this information is not read out but could in principle have been obtained (e.g.\ by detecting the escaping neutrino), the GSI experiment falls into category 2B and one has to sum over probabilities. This logic works because we know that the neutrino is, after some propagation, no superposition of mass eigenstates anymore, but just one particular eigenstate with a completely fixed mass.\\
A viewpoint closer to the amplitude formulation would be: If the neutrino finally interacts, it has to ``decide'' which mass eigenstate it has, even if it was a superposition of several mass eigenstates before. This is then equivalent to the image of having produced one particular mass eigenstate from the beginning on.

\section{\label{sec:amps} Amplitudes - probably the easiest language to use}

In this section, I use time-dependent amplitudes for the different basis states to describe another example, namely charged pion decay, which I compare then to neutrino oscillations (with referring to the actual situation in the GSI-experiment). The logical steps needed to understand the familiar example of pion decay are exactly the same as the ones needed to understand what is going on at GSI. This description is clear enough to account for very different situations and allows for an easy and nearly intuitive understanding of the various cases. Furthermore, it yields an a posteriori justification of the view used in the preceding section.

\subsection{\label{sec:pion} Charged pion decay}

It is well-known that a charged pion (e.g.\ $\pi^+$) can decay into either a positron in combination with an electron neutrino, or into the corresponding pair of $\mu$-like particles. Let us consider the case of a pure (and normalized) initial state pion $|\pi^+\rangle$. As this state evolves with time (and is not monitored), it will become a coherent superposition of the parent-state, as well as all possible daughter states:
\begin{equation}
 |\pi^+ (t)\rangle=\mathcal{A}_\pi (t) |\pi^+\rangle+ \mathcal{A}_\mu (t) |\mu^+ \nu_\mu\rangle+ \mathcal{A}_e (t)|e^+ \nu_e\rangle,
 \label{eq:pion_1}
\end{equation}
where all time-dependence is inside the partial amplitudes $\mathcal{A}_i$. Of course, this state has to be normalized correctly:
\begin{equation}
 |\mathcal{A}_\pi (t)|^2 + |\mathcal{A}_\mu (t)|^2 + |\mathcal{A}_e (t)|^2=1,
 \label{eq:pion_2}
\end{equation}
with $\mathcal{A}_\pi (0)=1$ and $\mathcal{A}_\mu (0)=\mathcal{A}_e (0)=0$. One can understand Eq.~\eqref{eq:pion_1} in the following way: The state at time $t$ is a coherent superposition of the basis states $\{ |\pi^+\rangle, |\mu^+ \nu_\mu\rangle, |e^+ \nu_e\rangle \}$ with time-dependent coefficients. Note that the basis states are orthogonal. The outcome of a certain measurement is some state $|\Psi\rangle$: All that a detector does is projecting on just this state $|\Psi\rangle$. Of course, different detectors will in general be described by projections on different $|\Psi\rangle$'s, which is a reflection of the influence of the process of measurement on the measurement itself. If one wants to know the probability for measuring that particular state, one has to calculate it according to the standard formula,
\begin{equation}
 P(\Psi)=|\langle \Psi|\pi^+ (t) \rangle|^2.
 \label{eq:pion_3}
\end{equation}
The question is what $|\Psi\rangle$ looks like. To make that clear, let us discuss several cases:
\begin{itemize}

\item The (trivial) case is that there has been no detection at all: Then we have gained no information. This means that the projected state is just the time-evolved state itself (we do not know anything except for the time passed since the experiment has started), and we get
\begin{equation}
 |\langle \Psi|\pi^+ (t) \rangle|^2=|\langle \pi^+ (t)|\pi^+ (t) \rangle|^2=1.
 \label{eq:pion_4}
\end{equation}
This result is trivial, since the probability for anything to happen must be equal to 1.

\item The next situation is when our experimental apparatus can give us the information that the pion has decayed, but we do not know the final state. Then, it can be either $|\mu^+ \nu_\mu\rangle$ or $|e^+ \nu_e\rangle$ and we remain with a superposition of these two states. The only information that we have gained is that the amplitude for the initial pion to be still there is now zero, $\mathcal{A}_\pi=0$ in Eq.~\eqref{eq:pion_1}. Then, the properly normalized state $|\Psi\rangle$ is
\begin{equation}
 | \Psi\rangle = \frac{\mathcal{A}_\mu (t) |\mu^+ \nu_\mu\rangle+ \mathcal{A}_e (t)|e^+ \nu_e\rangle}{\sqrt{|\mathcal{A}_\mu (t)|^2+|\mathcal{A}_e (t)|^2}}.
 \label{eq:pion_5}
\end{equation}
The absolute value square of the corresponding projection is
\begin{equation}
 |\langle \Psi|\pi^+ (t) \rangle|^2=|\mathcal{A}_\mu (t)|^2+|\mathcal{A}_e (t)|^2,
 \label{eq:pion_6}
\end{equation}
and if there is any oscillatory phase in the amplitudes, $\mathcal{A}_k (t)= \mathcal{\tilde A}_k (t) e^{i\omega_k t}$, it will have no effect due to the absolute values.

\item What if we know that the initial pion is still present? This sets $\mathcal{A}_\mu (t)=\mathcal{A}_e (t)=0$, and $|\Psi\rangle$ is just $\mathcal{A}_\pi (t) |\pi^+\rangle /\sqrt{|\mathcal{A}_\pi (t)|^2}$. The projection gives
\begin{equation}
 |\langle \Psi|\pi^+ (t) \rangle|^2=|\mathcal{A}_\pi (t)|^2,
 \label{eq:pion_7}
\end{equation}
which again does not oscillate.

\item If one particular final state, let us say $|e^+ \nu_e\rangle$, is detected, then $\mathcal{A}_\pi (t)=\mathcal{A}_\mu (t)=0$ and we get another term free of oscillations:
\begin{equation}
 |\langle \Psi|\pi^+ (t) \rangle|^2=|\mathcal{A}_e (t)|^2.
 \label{eq:pion_8}
\end{equation}

\end{itemize}
The question remains when we do get oscillations at all. The answer is: It depends on what our detector measures. If, e.g., the detector measures not exactly the state $|\mu^+ \nu_\mu\rangle$ or $|e^+ \nu_e\rangle$, but instead some (hypothetical) superposition (e.g., some quantum number which is not yet known, under which neither $\mu^+$ nor $e^+$ is an eigenstate, but some superposition of them), then one could measure the following (correctly normalized!) state:
\begin{equation}
 |\Psi\rangle=\frac{1}{\sqrt{2}} \left(|\mu^+ \nu_\mu\rangle + |e^+ \nu_e\rangle \right).
 \label{eq:pion_9}
\end{equation}
The squared overlap is
\begin{equation}
 |\langle \Psi|\pi^+ (t) \rangle|^2=\frac{1}{2}\left[ |\mathcal{A}_\mu (t)|^2+|\mathcal{A}_e (t)|^2 + 2 \Re \left( \mathcal{A}_\mu^* (t) \mathcal{A}_e (t) \right) \right],\nonumber
\end{equation}
where the $2 \Re \left( \mathcal{A}_\mu^* (t) \mathcal{A}_e (t) \right)$-piece will, in general, lead to oscillatory terms. What has been done differently than before? This time, we have done more than simply killing one or more amplitudes in Eq.~\eqref{eq:pion_1}, and this is the cause of oscillations: Whenever we are in a situation, in which the state playing the role of $|\Psi\rangle$ in Eq.~\eqref{eq:pion_9} is physical, the corresponding projection will yield oscillatory terms. As we will see in a moment, this is exactly what happens in neutrino oscillations.

\subsection{\label{sec:neutrino} Neutrino oscillations and the GSI-experiment}

Let us now turn to neutrino oscillations. Here, as we will see, a state like $|\Psi\rangle$ in Eq.~\eqref{eq:pion_9} can indeed be physical in some situations. To draw a clean analogy to the experiment done at GSI, we consider a hydrogen-like ion as initial state $|M\rangle$ that can decay to the state $|D \nu_e\rangle$ via electron capture. Since there was an electron in the initial state, we know that the amplitude for producing the mass eigenstate $|\nu_i\rangle$ is just $U_{ei}$. If there is no relative phase between the two mass eigenstates, the neutrino produced in the decay is exactly the particle that we call {\it electron neutrino}. In any case, due to different kinematics, the two mass eigenstates will in general develop different phases in the time-evolution. This means that, in spite of the mixing matrix elements $U_{ei}$ being time-independent, there will be a phase between the two neutrino mass eigenstates. Completely analogous to Eq.~\eqref{eq:pion_1}, the time-evolution of the initial state will be given by:
\begin{equation}
 | M(t)\rangle = \mathcal{A}_M (t) |M\rangle + U_{e1} \mathcal{A}_1 (t) |D \nu_1\rangle+U_{e2} \mathcal{A}_2 (t) |D \nu_2\rangle,
 \label{eq:neutrino_1}
\end{equation}
with $|\mathcal{A}_M (t)|^2+|U_{e1}\mathcal{A}_1 (t)|^2+|U_{e2}\mathcal{A}_2 (t)|^2=1$ and $\mathcal{A}_M (0)=1$.
We can immediately look at different cases:
\begin{itemize}

\item The parent ion is seen in the experiment: This kills all daughter amplitudes, $\mathcal{A}_{1,2} (t)=0$. The only remaining amplitude is $\mathcal{A}_M (t)$, very similar to Eq.~\eqref{eq:pion_7}. With the proper normalization for $|\Psi\rangle$ one gets no oscillation again:
\begin{equation}
 |\langle \Psi | M(t)\rangle |^2 = |\mathcal{A}_M (t)|^2
 \label{eq:neutrino_2}
\end{equation}

\item The next case corresponds to the GSI-experiment: One sees only the decay, but cannot tell which of the two neutrino mass eigenstates has been produced. This leads to $\mathcal{A}_M (t)=0$ and one has to perform a projection on the state
\begin{equation}
 |\Psi \rangle = \frac{U_{e1}\mathcal{A}_1 (t) |D \nu_1\rangle+U_{e2}\mathcal{A}_2 (t)|D \nu_2\rangle}{\sqrt{|U_{e1}\mathcal{A}_1 (t)|^2+|U_{e2}\mathcal{A}_2 (t)|^2}}.
 \label{eq:neutrino_3}
\end{equation}
Doing this with $|M(t)\rangle$ from Eq.~\eqref{eq:neutrino_1} yields
\begin{eqnarray}
 && |\langle \Psi | M(t)\rangle |^2= \left| \frac{|U_{e1}\mathcal{A}_1 (t)|^2 \cdot 1 +|U_{e2}\mathcal{A}_2 (t)|^2 \cdot 1}{\sqrt{|U_{e1}\mathcal{A}_1 (t)|^2+|U_{e2}\mathcal{A}_2 (t)|^2}} \right|^2=\nonumber\\
 && =|U_{e1}\mathcal{A}_1 (t)|^2+|U_{e2}\mathcal{A}_2 (t)|^2,
 \label{eq:neutrino_4}
\end{eqnarray}
which exhibits no oscillations, but is rather an incoherent sum over probabilities. This result is the justification of the intuitive treatment in Sec.~2.3: The elementary QM-discussion using probability amplitudes gives us just the correct prescription for how to sum up the amplitudes for the final states.

\item The GSI-experiment with infinite kinematical precision: In this case, one could actually distinguish the states $|D \nu_1\rangle$ and $|D \nu_2\rangle$. If one knows that $|D \nu_1\rangle$ is produced (e.g., by having very precise information about the kinematics), one will again have no oscillation,
\begin{equation}
 |\langle \Psi | M(t)\rangle |^2 = |\mathcal{A}_1 (t) U_{e1}|^2,
 \label{eq:neutrino_5}
\end{equation}
just as in Eq.~\eqref{eq:pion_8}.

\end{itemize}
These are in principle all cases that can appear. One can, however, have a closer look at the realistic situation in the GSI-experiment. Let us re-consider Eq.~\eqref{eq:neutrino_1}: In reality, the parent ion will be described by a wave packet with a finite size or, equivalently, a finite spreading in momentum space, due to the Heisenberg uncertainty relation. If this wave-packet is broad enough that each component can equivalently decay into $|D \nu_1\rangle$ or $|D \nu_2\rangle$, then both of the corresponding amplitudes will actually have the same phase ($\mathcal{A}_1 (t)=\mathcal{A}_2 (t)$), since they have the same energy, and one can write Eq.~\eqref{eq:neutrino_1} as
\begin{equation}
 | M(t)\rangle = \mathcal{A}_M (t) |M\rangle + \mathcal{A} (t)  \underbrace{\left[ U_{e1} |D \nu_1\rangle+U_{e2} |D \nu_2\rangle \right]}_{=|D \nu_e\rangle}.
 \label{eq:neutrino_6}
\end{equation}
Since the knowledge of the momentum of the parent ion is not accurate enough at the GSI-experiment to make a distinction between both final states $|D \nu_k\rangle$, this is a realistic situation. Of course, this does not at all change the above argumentation, since the final state $|\Psi\rangle$ will experience the same modification. The neutrino produced is an electron-neutrino, as to be expected.\\
The question remains, why some authors come to the conclusion that there should be oscillations? The answer is simple: If the correspondence between time-evolved initial state and detected state is wrong, then oscillations may appear. As example, we will consider the situation that the kinematics of the parent and daughter are fixed so tightly, that indeed the production amplitudes for $|D \nu_1\rangle$ and $|D \nu_2\rangle$ are not equal. This would correspond to an extremely narrow wave packet in momentum space. Let us, e.g, have in mind the extreme case when by kinematics only the production of $\nu_1$ is possible. This is no problem in principle and we would be used to it if neutrinos had higher masses, so that the $Q$-value of the capture was only sufficient to produce the lightest neutrino mass eigenstates. If only the disappearance of the parent is seen, the corresponding state $|\Psi\rangle$, which is detected, is given by Eq.~\eqref{eq:neutrino_3} (with $\mathcal{A}_2 (t)=0$ in the extreme case, but anyway with $\mathcal{A}_1 (t)\neq \mathcal{A}_2 (t)$). The corresponding neutrino is, however, no electron-neutrino anymore (which would be $U_{e1} |\nu_1\rangle+U_{e2} |\nu_2\rangle$, with the same phase for both states)! Indeed this is no surprise at all, since the kinematics in the situation considered is so tight that it changes the neutrino state which is emitted. This is a clear consequence of quantum mechanics, since for obtaining the necessary pre-knowledge (namely the very accurate information about the kinematics), one has to do a measurement that is precise enough to have an impact on the QM state.\\
If one consideres the state from Eq.~\eqref{eq:neutrino_3} as being the one emitted but then projects onto an electron neutrino state, oscillations will appear:
\begin{eqnarray}
 && |\langle D,\nu_e|M(t)\rangle|^2= |(U_{e1}^* \langle D \nu_1 |+U_{e2}^* \langle D \nu_2 |) \cdot \nonumber\\
 && \cdot(\mathcal{A}_M (t) |M\rangle + U_{e1} \mathcal{A}_1 (t) |D \nu_1\rangle+U_{e2} \mathcal{A}_2 (t) |D \nu_2\rangle)|^2=\nonumber\\
 && =|\mathcal{A}_1 (t)|^2+|\mathcal{A}_2 (t)|^2+2\Re (\mathcal{A}_1 (t) \mathcal{A}_2^* (t)).
 \label{eq:neutrino_7}
\end{eqnarray}
This is, however, wrong: One has not used all the information that could in principle have been obtained! But Nature does not care about if one uses information or not, so this treatment does simply not correspond to what has happened in the actual experiment. The oscillations, however, only arise due to the incorrect projection, and have no physical meaning.\\
The remaining question to obtain a complete understanding of the situation is if the neutrino that is emitted in the GSI-experiment oscillates. The answer is yes, of course. But to see that, we will have to modify our formalism a bit. Knowing that an electron neutrino has been emitted corresponds to $\mathcal{A}_M (t)=0$ in Eq.~\eqref{eq:neutrino_6}, and the remaining (normalized) state is:
\begin{equation}
 | \Psi\rangle = \frac{\mathcal{A} (t)}{|\mathcal{A} (t)|}\left[ U_{e1} |D \nu_1\rangle+U_{e2} |D \nu_2\rangle \right].
 \label{eq:neutrino_8}
\end{equation}
Re-phasing this state and measuring the time from $t$ on gives as initial state:
\begin{equation}
 | \Psi\rangle =U_{e1} |D \nu_1\rangle+U_{e2} |D \nu_2\rangle.
 \label{eq:neutrino_9}
\end{equation}
This is the state which will undergo some evolution in time according to
\begin{equation}
 | \Psi (t')\rangle =\mathcal{A}'_1(t') U_{e1} |D \nu_1\rangle+\mathcal{A}'_2(t')U_{e2} |D \nu_2\rangle,
 \label{eq:neutrino_10}
\end{equation}
with $|\mathcal{A}'_1(t') U_{e1}|^2+|\mathcal{A}'_2(t') U_{e2}|^2=1$ and $\mathcal{A}'_1(0)=\mathcal{A}'_2(0)=1$. If we ask what happens to this neutrino if it is detected after some macroscopic distance, it is necessary to take into account what has happend to the daughter nucleus that has been produced together with the neutrino, due to entanglement. The daughter nucleus, which is accurately described by a wave packet, is detected, but not with sufficient kinematical accuracy to distinguish the different components $|D\rangle$ of the wave packet. The effect of such a non-measurement is studied most easily in the density matrix formalism. The density matrix $\rho'$ corresponding to Eq.~\eqref{eq:neutrino_10} is given by
\begin{eqnarray}
 && |\Psi (t')\rangle \langle \Psi(t')| =|\mathcal{B}_1(t')|^2 |D\rangle |\nu_1\rangle \langle \nu_1| \langle  D|+\nonumber\\
 && +|\mathcal{B}_2(t')|^2 |D\rangle |\nu_2\rangle \langle \nu_2| \langle  D|+\nonumber\\
 &&+[\mathcal{B}_1(t') \mathcal{B}^*_2(t') |D\rangle |\nu_1\rangle  \langle \nu_2| \langle  D|+h.c.],
 \label{eq:neutrino_11}
\end{eqnarray}
where $\mathcal{B}_k(t')=\mathcal{A}'_k(t') U_{ek}$. If the exact kinematics of the daughter is not measured, then one has to calculate the trace over the corresponding states. It gives
\begin{eqnarray}
 &&\rho\equiv \int dD \langle D|\rho'|D\rangle=|\mathcal{B}_1(t')|^2 |\nu_1\rangle \langle \nu_1| +\label{eq:neutrino_12}\\
 &&+|\mathcal{B}_2(t')|^2 |\nu_2\rangle \langle \nu_2|+ (\mathcal{B}_1(t') \mathcal{B}^*_2(t') |\nu_1\rangle \langle \nu_2| +h.c.).\nonumber
\end{eqnarray}
If we want to know the probability to detect, e.g., a muon neutrino, $|\nu_\mu\rangle=U_{\mu 1} |\nu_1\rangle+ U_{\mu 2}|\nu_2\rangle$, the corresponding projection operator is given by
\begin{equation}
 \mathcal{P}_\mu= |\nu_\mu\rangle \langle \nu_\mu|,
 \label{eq:neutrino_13}
\end{equation}
and the probability to detect this state is
\begin{equation}
 P_\mu={\rm Tr}(\mathcal{P}_\mu \rho)=\langle \nu_1|\mathcal{P}_\mu \rho|\nu_1\rangle+\langle \nu_2|\mathcal{P}_\mu \rho|\nu_2\rangle.
 \label{eq:neutrino_14}
\end{equation}
Note, however, that the neutrino states $|\nu_{1,2}\rangle$ will always be orthogonal, since they correspond to eigenstates of different masses (like an electron is in that sense orthogonal to a muon). The result is
\begin{eqnarray}
 && P_\mu= |U_{\mu 1}|^2 |\mathcal{B}_1(t')|^2 + |U_{\mu 2}|^2 |\mathcal{B}_2(t')|^2+\nonumber\\
 && +[U_{\mu 1} U_{\mu 2}^* \mathcal{B}^*_1(t') \mathcal{B}_2(t') +c.c.],
 \label{eq:neutrino_15}
\end{eqnarray}
whose second line contains oscillatory contributions. These oscillation are indeed physical: Eq.~\eqref{eq:neutrino_13} is a description of a detector that is sensitive to $\nu_\mu$'s only. If it could not distinguish different neutrino flavours, the oscillation would vanish again.

\section{\label{sec:beats} Quantum Beats}
\begin{figure}[t]
  \begin{center}
   \includegraphics[height=5.2cm]{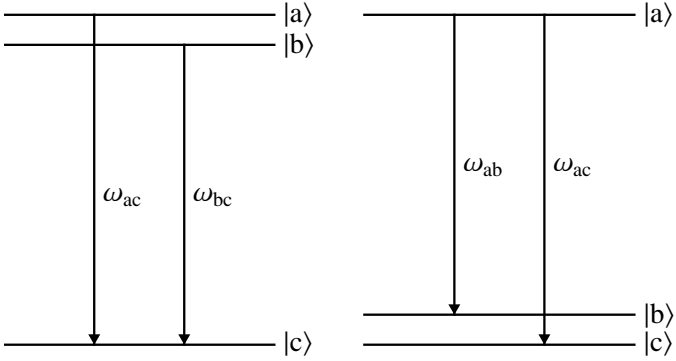}
  \end{center}
  \caption{Type I (left) and type II (right) of the Quantum Beats settings.}
  \label{fig:QBs}
\end{figure}

The last point to discuss are the so-called Quantum Beats (QBs)~\cite{Chow:1975zz}. This phenomenon is known from Quantum Optics and has often been mentioned as possible explanation for the GSI anomaly. As we will see, the corresponding language can be equally used to describe the GSI-experiment and (of course) yields the same result as already obtained. Still, it is also useful to consider the experiment from this point of view in order not to be misled by claims that erroneously make QBs arising from a splitting in the final state responsible for the observation at GSI.\\
Normally, one considers atomic levels for this discussion, and we will stick to that here for illustrative purposes and give the relation to the GSI-experiment at the end of each section. This way also easily clarifies the analogies to the Quantum Optics formulation.

\subsection{\label{sec:beats_typeI} Single atom of type I}

Let us start with the classic example of QBs, namely an atom in a coherent superposition of three states $|a\rangle$, $|b\rangle$, and $|c\rangle$, where the first two states are above and closely spaced compared to $|c\rangle$. This setting is drawn on the left panel of Fig.~\ref{fig:QBs} and is referred to as ``type~I''. First note that the three levels correspond to different (but fixed) eigenvalues of the energy and are hence orthogonal vectors in Hilbert space. This is not at all changed by an energy uncertainty which, however, makes it possible to have a coherent superposition of the three states. Initially, we assume the atom to be in such a superposition of these states, but having emitted no photon yet. Accordingly, the photon state can only be the vacuum $|0\rangle_\gamma$. Then, the initial state of this system can be written as
\begin{equation}
 |\Psi(0)\rangle=\mathcal{A}_0|a\rangle |0\rangle_\gamma+\mathcal{B}_0|b\rangle |0\rangle_\gamma+\mathcal{C}_0|c\rangle |0\rangle_\gamma,
 \label{eq:QBs_1}
\end{equation}
where $|\mathcal{A}_0|^2+|\mathcal{B}_0|^2+|\mathcal{C}_0|^2=1$. If this system undergoes a time-evolution, the lower state might be populated by de-excitation of the upper ones, which is done by photon emission. If the state $|1_x\rangle_\gamma=a^\dagger_x |0\rangle_\gamma$ is assumed to describe a state with one photon of frequency $\om_x$, then the state at time $t$ can be written as
\begin{eqnarray}
 && |\Psi(t)\rangle=\mathcal{A}(t)|a\rangle |0\rangle_\gamma+\mathcal{B}(t)|b\rangle |0\rangle_\gamma+\mathcal{C}(t)|c\rangle |0\rangle_\gamma+\nonumber\\
 && + \mathcal{C}_1(t)|c\rangle |1_{ac}\rangle_\gamma + \mathcal{C}_2(t)|c\rangle |1_{bc}\rangle_\gamma,
 \label{eq:QBs_2}
\end{eqnarray}
where $\mathcal{A}(0)=\mathcal{A}_0$, $\mathcal{B}(0)=\mathcal{B}_0$, $\mathcal{C}(0)=\mathcal{C}_0$, $\mathcal{C}_{1,2}(0)=0$, and $|\mathcal{A}(t)|^2+|\mathcal{B}(t)|^2+|\mathcal{C}(t)|^2+|\mathcal{C}_1(t)|^2+|\mathcal{C}_2(t)|^2=1$. Under the assumption that all levels are equally populated, the radiated intensity will be proportional to $\langle \Psi(t)| \vec{E}^2(\vec{0},t) |\Psi(t)\rangle$, where
\begin{equation}
 \vec{E}(\vec{x},t)=\sum_{\vec{k},\lambda} \eps_{\vec{k},\lambda} \left( a_{\vec{k},\lambda} e^{-ikx} + a_{\vec{k},\lambda}^\dagger e^{+ikx} \right)
 \label{eq:QBs_3}
\end{equation}
is the electric field operator and $\eps_{\vec{k},\lambda}$ is the electric
field per photon of momentum $\vec{k}$ and polarization $\lambda$. Note that the
creation and annihilation operators have only one non-trivial commutation
relation, namely
$[a_{\vec{k},\lambda},a_{\vec{k}',\lambda'}^\dagger]=\delta_{\vec{k},\vec{k}'} \delta_{\lambda,\lambda'}$. In our case we obtain effectively:
\begin{eqnarray}
 &&\vec{E}(\vec{0},t)^2=\eps_{ac}^2 (1+2a_{ac}^\dagger a_{ac})+\label{eq:QBs_4}\\
 &&+\eps_{bc}^2 (1+2a_{bc}^\dagger a_{bc})+ 2 \eps_{ac} \eps_{bc} (a_{ac}^\dagger a_{bc} e^{i\Delta t} + a_{bc}^\dagger a_{ac} e^{-i\Delta t}),\nonumber
\end{eqnarray}
where $\Delta=\om_{ac}-\om_{bc}$. Here, we have already used that terms like, e.g., $a_{ac}^2$ give no contribution with $|\Psi\rangle$ from Eq.~\eqref{eq:QBs_2}. Remember now, that the atomic states are orthonormal. This means that one can, e.g., combine a term proportional to $\langle b|$ in $\langle \Psi(t)|$ only with the corresponding term $|b\rangle$ in $|\Psi(t)\rangle$. The corresponding combination of amplitudes $|\mathcal{B}(t)|^2$ does, however, not oscillate, since any phase will be killed by the absolute value. This is also true for every term involving one of the constant parts of Eq.~\eqref{eq:QBs_4}: E.g.\ the term proportional to $\mathcal{C}^*(t) \mathcal{C}_1(t)$ can involve a factor
\begin{equation}
 {_\gamma} \langle 0| a_{ac}^\dagger a_{ac} a_{ac}^\dagger |0\rangle_\gamma= 0,
 \label{eq:QBs_5}
\end{equation}
because of $a_{ac}^\dagger$ acting on the left. There are, however, remaining oscillatory terms such as $\mathcal{C}_1^*(t) \mathcal{C}_2(t) e^{i\Delta t}$, which is proportional to
\begin{equation}
 {_\gamma} \langle 0|a_{ac} a_{ac}^\dagger a_{bc} a_{bc}^\dagger |0\rangle_\gamma= {_\gamma} \langle 0|(1+a_{ac}^\dagger a_{ac}) (1+a_{bc}^\dagger a_{bc}) |0\rangle_\gamma=1.\nonumber
\end{equation}
These terms cause the Quantum Beats for a type I atom. Actually, one could have expected this result intuitively: Both of the coherently excited upper levels can decay into {\it the same} state $|c\rangle$ via the emission of a photon. Hence, one cannot in any way determine the photon energy without measuring it directly. Without such a measurement, interference terms will appear.

How is the situation for the GSI-experiment? In this case one simply has to replace the photon by the neutrino. As explained in Ref.~\cite{Kienert:2008nz} for instance, a splitting in the initial state could lead to an oscillatory behavior. This splitting, however, would have to be tiny, $\sim 10^{-15}$~eV, a value which can hardly be explained. Furthermore, there exists preliminary data on the lifetimes of $^{142}{\rm Pm}^{60+}$ with respect to $\beta^+$-decay that shows no oscillatory behavior~\cite{Ivanov:2009rc}. An initial splitting in the nucleus would lead to an oscillatory rate in this case, too. Accordingly, if such a splitting is present in the initial state, it could be in the levels of the single bound electron, since this would then affect EC-decays while leaving $\beta^+$-decays untouched.

\subsection{\label{sec:beats_typeII} Single atom of type II}

We can study a similar setting, namely an atom of type~II, shown on the right panel of Fig.~\ref{fig:QBs}. The corresponding initial state would again be described by Eq.~\eqref{eq:QBs_1}, but its time-evolution would now look like
\begin{eqnarray}
 && |\Psi(t)\rangle=\mathcal{A}(t)|a\rangle |0\rangle_\gamma+\mathcal{B}(t)|b\rangle |0\rangle_\gamma+\mathcal{C}(t)|c\rangle |0\rangle_\gamma+\nonumber\\
 && + \mathcal{B}'(t)|b\rangle |1_{ab}\rangle_\gamma + \mathcal{C}'(t)|c\rangle |1_{ac}\rangle_\gamma,
 \label{eq:QBs_7}
\end{eqnarray}
where $\mathcal{A}(0)=\mathcal{A}_0$, $\mathcal{B}(0)=\mathcal{B}_0$, $\mathcal{C}(0)=\mathcal{C}_0$, $\mathcal{B}'(0)=0$, $\mathcal{C}'(0)=0$, and $|\mathcal{A}(t)|^2+|\mathcal{B}(t)|^2+|\mathcal{C}(t)|^2+|\mathcal{B}'1(t)|^2+|\mathcal{C}'(t)|^2=1$. The square of the electric field has again the form of Eq.~\eqref{eq:QBs_4}, just with $bc \rightarrow ab$. Due to the orthogonality of the atomic states, there are not too many combinations which are possible:
\begin{itemize}

\item 0-photon state coupled with itself:\\
If we take, e.g., the term $|\mathcal{A}(t)|^2$, it does not oscillate anyway. Hence, only the time-dependent parts in Eq.~\eqref{eq:QBs_4} (with $bc \rightarrow ab$) could lead to oscillations. But they are proportional to
\begin{equation}
 {_\gamma} \langle 0| a_{ac}^\dagger a_{ab} |0\rangle_\gamma={_\gamma} \langle 0| a_{ab}^\dagger a_{ac} |0\rangle_\gamma=0.\nonumber
\end{equation}

\item 1-photon state coupled with itself:\\
$|\mathcal{B}'(t)|^2$ does not oscillate, too, and the time-dependent terms from the electric field yield
\begin{eqnarray}
 && {_\gamma} \langle 1_{ab}| a_{ac}^\dagger a_{ab} |1_{ab}\rangle_\gamma={_\gamma} \langle 0| a_{ab} a_{ac}^\dagger a_{ab} a_{ab}^\dagger|0\rangle_\gamma=0\ {\rm and}\nonumber\\
 && {_\gamma} \langle 1_{ab}| a_{ab}^\dagger a_{ac} |1_{ab}\rangle_\gamma={_\gamma} \langle 0| a_{ab} a_{ab}^\dagger a_{ac} a_{ab}^\dagger|0\rangle_\gamma=0, \nonumber
\end{eqnarray}
which follows immediately from the action of $a_{ac}^\dagger$ to the left and of $a_{ac}$ to the right, respectively.

\item  0-photon state coupled with 1-photon state:\\
This is the only possibility, which is left. If we take for instance the term $\mathcal{B}^*(t) \mathcal{B}'(t)$, this will oscillate anyway, so we will also have to check the constant terms in Eq.~\eqref{eq:QBs_4}. The ones proportional to $1$ are naturally zero, ${_\gamma} \langle 0|1_{ab}\rangle_\gamma={_\gamma} \langle 0| a_{ab}^\dagger |0\rangle_\gamma=0.$ The other terms are
\begin{eqnarray}
 && {_\gamma} \langle 0| \underbrace{a_{ac}^\dagger}_{0\leftarrow} a_{ac} |1_{ab}\rangle_\gamma=0,\ {_\gamma} \langle 0| \underbrace{a_{ab}^\dagger}_{0\leftarrow} a_{ab} |1_{ab}\rangle_\gamma=0,\label{eq:QBs_8}\\
 && {_\gamma} \langle 0| \underbrace{a_{ac}^\dagger}_{0\leftarrow} a_{ab} |1_{ab}\rangle_\gamma=0,\ {\rm and}\ {_\gamma} \langle 0| \underbrace{a_{ab}^\dagger}_{0\leftarrow} a_{ac} |1_{ab}\rangle_\gamma=0,\nonumber
\end{eqnarray}
where the action of the operators to give zero is always indicated by the arrow. The argumentation is analogous for the complex conjugated term.

\end{itemize}
Hence, there can be no Quantum Beats for a single atom of type II! The intuitive reason is that, by waiting long enough, one could reach an accuracy in energy that is good enough to distinguish the possible final states $|b\rangle$ and $|c\rangle$. This would then be a way to determine the energy of the emitted photon without disturbing it.

To give an analogous argument for the GSI-experiment, one has to turn the comparison given in Sec.~4.1 round and replace the {\it atom by the neutrino} and the {\it photon by the ion}. The reason is that what is claimed to interfere in this situation is the neutrino states themselves (see, e.g., Ref.~\cite{Ivanov:2008sd}). This neutrino is not expected to interact, before losing its coherence (cf.\ Sec.~2.3). However, once it interacts, it has to decide for a certain mass eigenstate. By monitoring this interaction, it would in principle be no problem to determine the neutrino's mass (e.g., by exploiting the spatial separation of the mass eigenstates far away from the source) and from this one could easily reconstruct the kinematics of the daughter ion in the GSI-experiment. Accordingly, no QBs are to be expected in this situation.

\subsection{\label{sec:beats_twotypeII} Two atoms of type II}

On the other hand, there is a situation in which we can expect QBs even for atoms of type II, namely if we have two of them. If these two atoms are separated by a distance which is smaller than the wavelength of the emitted photons, there is no way to resolve their separation in space and we have to write down a combined initial state for both atoms, 1 and 2:
\begin{eqnarray}
 && |\Psi(0)\rangle=\mathcal{A}_0 |a\rangle_1 |a\rangle_2 |0\rangle_\gamma+\mathcal{B}_0 |b\rangle_1 |b\rangle_2 |0\rangle_\gamma+\mathcal{C}_0 |c\rangle_1 |c\rangle_2 |0\rangle_\gamma\nonumber\\
 && +\mathcal{D}_{1,0} |a\rangle_1 |b\rangle_2 |0\rangle_\gamma+\mathcal{D}_{2,0} |b\rangle_1 |a\rangle_2 |0\rangle_\gamma+\mathcal{E}_{1,0} |a\rangle_1 |c\rangle_2 |0\rangle_\gamma+\nonumber\\
 && +\mathcal{E}_{2,0} |c\rangle_1 |a\rangle_2 |0\rangle_\gamma+\mathcal{F}_{1,0} |b\rangle_1 |c\rangle_2 |0\rangle_\gamma+\mathcal{F}_{2,0} |c\rangle_1 |b\rangle_2 |0\rangle_\gamma.\nonumber
\end{eqnarray}
The corresponding time-evolution $|\Psi(t)\rangle$ looks a bit complicated:
\begin{eqnarray}
 && \mathcal{A}(t) |a\rangle_1 |a\rangle_2 |0\rangle_\gamma
+\mathcal{B}(t) |b\rangle_1 |b\rangle_2 |0\rangle_\gamma
+\mathcal{C}(t) |c\rangle_1 |c\rangle_2 |0\rangle_\gamma+\nonumber\\
 && +\mathcal{D}_1(t) |a\rangle_1 |b\rangle_2 |0\rangle_\gamma
+\mathcal{D}_2(t) |b\rangle_1 |a\rangle_2 |0\rangle_\gamma+\nonumber\\
 && +\mathcal{E}_1(t) |a\rangle_1 |c\rangle_2 |0\rangle_\gamma
+\mathcal{E}_2(t) |c\rangle_1 |a\rangle_2 |0\rangle_\gamma+\nonumber\\
 && +\mathcal{F}_1(t) |b\rangle_1 |c\rangle_2 |0\rangle_\gamma
+\mathcal{F}_2(t) |c\rangle_1 |b\rangle_2 |0\rangle_\gamma+\nonumber\\
 && +\mathcal{G}_1(t) |b\rangle_1 |a\rangle_2 |1_{ab}\rangle_\gamma
+\mathcal{G}_2(t) |a\rangle_1 |b\rangle_2 |1_{ab}\rangle_\gamma+\nonumber\\
 && +\mathcal{H}_1(t) |c\rangle_1 |a\rangle_2 |1_{ac}\rangle_\gamma
+\mathcal{H}_2(t) |a\rangle_1 |c\rangle_2 |1_{ac}\rangle_\gamma+\nonumber\\
 && +\mathcal{I}_1(t) |b\rangle_1 |b\rangle_2 |1_{ab}\rangle_\gamma
+\mathcal{I}_2(t) |c\rangle_1 |c\rangle_2 |1_{ac}\rangle_\gamma+\nonumber\\
 && +\mathcal{J}_1(t) |b\rangle_1 |c\rangle_2 |1_{ab}\rangle_\gamma
+\mathcal{J}_2(t) |c\rangle_1 |b\rangle_2 |1_{ab}\rangle_\gamma+\nonumber\\
 && +\mathcal{K}_1(t) |b\rangle_1 |c\rangle_2 |1_{ac}\rangle_\gamma
+\mathcal{K}_2(t) |c\rangle_1 |b\rangle_2 |1_{ac}\rangle_\gamma.
 \label{eq:QBs_9}
\end{eqnarray}
One oscillatory term would then be, e.g., $\mathcal{J}_1^* \mathcal{K}_1 e^{-i\Delta t}$, which is proportional to
\begin{eqnarray}
 && {_\gamma} \langle 1_{ab}| a_{ab}^\dagger a_{ac} |1_{ac}\rangle_\gamma={_\gamma} \langle 0| a_{ab} a_{ab}^\dagger a_{ac} a_{ac}^\dagger |0\rangle_\gamma=\label{eq:QBs_10}\\
 && ={_\gamma} \langle 0| (1+\underbrace{a_{ab}^\dagger}_{0\leftarrow} a_{ab}) (1+a_{ac}^\dagger \underbrace{a_{ac}}_{\rightarrow 0}) |0\rangle_\gamma={_\gamma} \langle 0|0\rangle_\gamma=1.\nonumber
\end{eqnarray}
If the spatial separation is less than the photon wavelength, one cannot determine the photon energy, because one does not know which atom has emitted the radiation. Accordingly, we expect QBs.

For the GSI-case, this possibility has to be taken into account, because even for runs with one single EC only, there can have been more ions in the ring that were lost or decayed via $\beta^+$. In this case (comparing the neutrino again with the photon), one has to replace the wavelength of the photon by the de Broglie wavelength of the neutrino. The neutrino energy should be of the same order as the $Q$-value of the EC-reaction, which is roughly 1~MeV~\cite{Litvinov:2008rk}. The corresponding wavelength is, however, $\lambda=\frac{2\pi\hbar c}{Ec}\sim 10^{-12}$~m, while the average distance between two ions should be of the order of the storage ring~\cite{Steck:1996me}, which is roughly 100~m~\cite{ESR}. Hence, this possibility is excluded for the GSI-experiment.

\section{\label{sec:conclusions} Conclusions}

A comparison of the GSI-experiment with several other processes (the Double Slit experiment with photons, $e^+ e^- \rightarrow \mu^+ \mu^-$ scattering, and charged pion decay) has been given. By using the language of QFT as well as the intuitive formulation with probability amplitudes, I have shown that the situation at GSI cannot lead to any oscillation of the decay rate, if the correct treatment is chosen and no additional assumptions (as, e.g., a splitting in the initial state) are taken into account. Also the frequently mentioned possibility of Quantum Beats of the final state cannot explain the observed oscillations, at least not in the standard picture. Hopefully this article will contribute to the clarification of the physical situation in the experiment that has been performed at GSI.

\section*{\label{sec:ack} Acknowledgements}

I would like to thank A.~Blum and M.~Lindner for useful discussions, as well as K.~Blaum and especially J.~Kopp for carefully reading the manuscript and giving valuable comments. I am furthermore grateful to the referee of this paper for making valuable suggestions that have improved the manuscript considerably. This work has been supported by Deutsche Forschungsgemeinschaft in the Sonderforschungsbereich Transregio~27.

\bibliographystyle{./apsrev}
\bibliography{./GSI-DSlit-refs}

\end{document}